\documentclass[conference]{IEEEtran}
\IEEEoverridecommandlockouts
\usepackage{cite}
\usepackage{amsmath,amssymb,amsfonts}
\usepackage{algorithmic}
\usepackage{graphicx}
\usepackage{textcomp}
\usepackage{xcolor}
\usepackage{multirow}
\def\BibTeX{{\rm B\kern-.05em{\sc i\kern-.025em b}\kern-.08em
    T\kern-.1667em\lower.7ex\hbox{E}\kern-.125emX}}
\begin{document}

\title{WaggleNet: A LoRa and MQTT-Based Monitoring System for Internal and External Beehive Conditions}

\author{
\centering
\begin{tabular}{ccc}
\textbf{1\textsuperscript{st} Minju Jeon} & \textbf{2\textsuperscript{nd} Jiyun Kim} & \textbf{3\textsuperscript{rd} Sewon Kim} \\
\textit{AI Computer Science and Engineering} & \textit{College of Information Science} & \textit{Computer Science and Engineering} \\
Kyonggi University & Hallym University & Jeonbuk National University \\
Suwon, South Korea & Chuncheon, South Korea & Jeonju, South Korea \\
mingmingmon@kyonggi.ac.kr & 20225141@hallym.ac.kr & andimsewon@jbnu.ac.kr \\
\\[-0.6em]
\textbf{4\textsuperscript{th} Seongmin Park} & \textbf{5\textsuperscript{th} Bo Zhang} & \textbf{6\textsuperscript{th} Anthony H. Smith} \\
\textit{Computer Science and Engineering} & \textit{Polytechnic Institute} & \textit{Computer and Information Technology} \\
Kyonggi University & Purdue University & Purdue University \\
Suwon, South Korea & Indiana, United States & West Lafayette, IN, USA \\
duke7272@kyonggi.ac.kr & zhan5055@purdue.edu & ahsmith@purdue.edu \\
\end{tabular}
}

\maketitle

\begin{abstract}
Bee populations are declining globally due to habitat loss, pesticide exposure, and climate change, threatening agricultural productivity and food security. While existing smart beehive systems monitor internal conditions, they typically overlook external environmental factors that significantly influence colony health, and are constrained by high cost, limited scalability, and inadequate contextual analysis. We present WaggleNet, a novel dual-scope monitoring system that simultaneously captures both internal hive conditions and external environmental parameters using a cost-effective LoRa-MQTT architecture. Our system deploys modular worker nodes ($\sim$\$15 each) equipped with temperature, humidity, light, and GPS sensors both inside and around beehives. A master node functions as a LoRa-MQTT gateway, forwarding data to a cloud server with a mobile application interface. Field experiments confirmed reliable operation with 100\% packet delivery over 110 meters in line-of-sight conditions and 95 meters in obstructed environments, including successful deployment inside wooden hive structures. Our system demonstrated stable end-to-end latency under 5 seconds and continuous operation over a two-month period across diverse environmental conditions. By bridging the gap between internal and external monitoring, WaggleNet enables contextual anomaly detection and supports data-driven precision beekeeping in resource-constrained settings.
\end{abstract}


\section{Introduction}

Honeybees (Apis mellifera) contribute to the pollination of approximately 30\% of global food crops, providing ecosystem services valued at over \$200 billion annually \cite{kamboj2024understanding}. However, colony collapse disorder (CCD) and environmental stressors have led to alarming declines in bee populations worldwide. Recent studies demonstrate that environmental fluctuations in temperature, humidity, and light significantly disrupt bee behavior, colony thermoregulation, and brood development \cite{kim2024exposure, stabentheiner2010honeybee}. To address these threats, precision beekeeping—leveraging real-time sensor networks for proactive hive management—has emerged as a critical research direction.

Existing smart beehive monitoring systems have made substantial progress in digitizing apiculture. Commercial and research systems such as Arnia \cite{commercial_arnia}, Beemon \cite{bortolotti2021beemon}, and b+WSN \cite{edwards2016bwsn} employ wireless sensor networks to track internal hive parameters including temperature, humidity, weight, and acoustic signatures. However, these systems face four key limitations that constrain their real-world deployment and effectiveness:

\textbf{High Implementation Cost:} Commercial systems such as Arnia and SolutionBee cost \$200--500 per hive, creating economic barriers for small-scale beekeepers and academic researchers. Even research prototypes often rely on expensive components such as GSM/LTE modules (\$30--80) and specialized load cells (\$50--150), limiting scalability \cite{ntawuzumunsi2021sbmacs}.

\textbf{Limited Scalability:} Many systems use point-to-point WiFi or Bluetooth communication, requiring hives to be located within 50--100 meters of internet infrastructure. This constraint renders them impractical for distributed apiaries across large geographic areas \cite{ochoa2019iot}.

\textbf{Narrow Monitoring Scope:} Existing literature predominantly focuses on \textit{internal} hive conditions, overlooking the critical influence of \textit{external} environmental factors. Studies show that external ambient temperature, solar radiation, and weather patterns directly affect foraging behavior, thermoregulation effort, and colony productivity \cite{leach2024external}. The absence of external monitoring prevents accurate interpretation of internal anomalies and limits contextual decision-making.

\textbf{Inadequate Real-time Response:} Several systems store data for offline analysis or batch processing, lacking real-time alerting mechanisms that enable timely intervention during critical events such as thermal stress, predator intrusion, or environmental hazards \cite{murphy2015bwsn}.

To address these limitations, we present \textbf{WaggleNet}, a comprehensive monitoring system with the following contributions:

\begin{itemize}
\item \textbf{Dual-Scope Environmental Monitoring:} Unlike prior work that focuses exclusively on internal conditions, WaggleNet simultaneously monitors both internal hive parameters and external ambient conditions, enabling contextual interpretation of anomalies and improved decision-making.

\item \textbf{Cost-Effective Modular Design:} Using low-cost ESP32-based LoRa modules (\$10--15) and commodity sensors (DHT22, LDR, NEO-6M GPS), we achieve a total node cost of approximately \$25--30, representing an 80--85\% cost reduction compared to commercial alternatives while maintaining comparable functionality.

\item \textbf{Long-Range Scalable Architecture:} By integrating LoRa (long-range, low-power wireless) with MQTT (efficient publish-subscribe messaging), our system supports distributed deployments over 100+ meter ranges without requiring continuous WiFi coverage, enabling wide-area apiary monitoring.

\item \textbf{Real-Time Cloud-Mobile Integration:} A full-stack implementation including cloud storage (SQLite database), RESTful APIs, and a cross-platform mobile application (Flutter) provides real-time visualization, GPS-based spatial mapping, and push notification alerts for critical events.

\item \textbf{Field-Validated Reliability:} Two-month field trials in actual beekeeping environments confirm system robustness, with 100\% packet delivery in line-of-sight conditions, stable operation inside wooden hive structures, and end-to-end latency under 5 seconds for real-time monitoring.
\end{itemize}

The remainder of this paper is organized as follows: Section II reviews related work and positions our contributions; Section III describes system requirements; Section IV details the hardware-software architecture; Section V presents experimental evaluation including field deployments; Section VI discusses implications and limitations; and Section VII concludes with future directions.

\section{Related Work}

\subsection{Smart Beehive Monitoring Systems}

Recent systematic reviews \cite{zacepins2015systematic, iot_beehive_review2024} identify three primary categories of beehive monitoring systems: (1) environmental monitoring, (2) behavioral analysis, and (3) health diagnostics. Table \ref{tab:related_work} provides a quantitative comparison of representative systems.

\textbf{Environmental Monitoring Systems:} Edwards-Murphy et al. \cite{edwards2016bwsn} developed b+WSN, a heterogeneous wireless sensor network monitoring temperature, humidity, CO$_2$, and acoustic signatures inside beehives. While comprehensive in internal sensing, their system uses ZigBee communication (50m range) and lacks external environmental correlation. Zacepins et al. \cite{zacepins2015beekeeping} deployed temperature sensors across multiple hive locations but did not integrate external weather monitoring or real-time alerting.

Ntawuzumunsi and Kumaran \cite{ntawuzumunsi2021sbmacs} proposed SBMaCS using LoRa communication for remote monitoring, achieving 2 km range in rural settings. However, their system costs approximately \$120 per node due to expensive GSM modules and focuses solely on internal parameters. Bortolotti et al. \cite{bortolotti2021beemon} developed Beemon, a scalable IoT platform, but deployment requires WiFi infrastructure and does not capture external environmental context.

\textbf{Behavioral and Health Analysis:} Gil-Lebrero et al. \cite{gillebrero2017honey} used audio analysis and machine learning to detect queen presence and swarming events, achieving 89\% classification accuracy. However, their system requires high sampling rates (44.1 kHz audio) and significant computational resources. Kulyukin et al. \cite{kulyukin2018audio} demonstrated acoustic detection of bee traffic patterns but did not integrate environmental sensors for contextual analysis.

Recent work by Liyanage et al. \cite{liyanage2024iot} and Nayyar et al. \cite{nayyar2025hivelink} implemented IoT-based systems using NodeMCU and cloud platforms (ThingSpeak, Firebase). While these provide web dashboards, they lack: (1) external environmental monitoring, (2) GPS-based spatial mapping, and (3) systematic evaluation in real apiary deployments beyond laboratory prototypes.

\textbf{Commercial Systems:} Arnia Hive Monitoring and SolutionBee offer complete solutions with weight sensors, temperature probes, and cellular connectivity. However, their subscription-based models (\$200--500/year per hive) and proprietary designs limit adoption for small-scale beekeepers and research applications \cite{commercial_review}.

\subsection{LoRa-Based Agricultural IoT}

LoRa (Long Range) technology has gained traction in agricultural IoT due to its combination of long-range communication (2--10 km), low power consumption (10--100 mW), and unlicensed spectrum operation \cite{adelantado2017understanding}. Ochoa et al. \cite{ochoa2019iot} demonstrated a LoRa-based beehive system using Ubidots cloud platform, but their implementation lacks: (1) systematic range evaluation, (2) external environmental monitoring, and (3) GPS integration for multi-hive deployments.

Vatskel et al. \cite{vatskel2024energy} proposed an energy-efficient LoRa system for apiary management with modular programmable logic controllers, focusing on power optimization but not addressing the internal-external monitoring gap. Rahmatullah et al. \cite{rahmatullah2025multinode} analyzed LoRa-MQTT hybrid architectures and reported 30\% higher throughput compared to LoRaWAN in short-range scenarios (<5 km), validating our architectural choice.

Zhang et al. \cite{zhang2023iot} surveyed IoT implementations in underground mining and demonstrated LoRa's reliability in signal-attenuated environments, with successful transmission through concrete and soil barriers—insights applicable to beehive structures.

\begin{table*}[t]
\caption{Quantitative Comparison of Beehive Monitoring Systems}
\begin{center}
\small
\begin{tabular}{|l|c|c|c|c|c|c|c|}
\hline
\textbf{System} & \textbf{Cost/Node} & \textbf{Range} & \textbf{Internal} & \textbf{External} & \textbf{GPS} & \textbf{Real-time} & \textbf{Field} \\
& \textbf{(USD)} & \textbf{(m)} & \textbf{Monitor} & \textbf{Monitor} & \textbf{Tracking} & \textbf{Alerts} & \textbf{Validated} \\
\hline
b+WSN \cite{edwards2016bwsn} & \$80--100 & 50 & \checkmark & \texttimes & \texttimes & \texttimes & \checkmark \\
Beemon \cite{bortolotti2021beemon} & \$60--80 & 100 (WiFi) & \checkmark & \texttimes & \texttimes & \checkmark & \texttimes \\
SBMaCS \cite{ntawuzumunsi2021sbmacs} & \$120--150 & 2000 & \checkmark & \texttimes & \texttimes & \checkmark & \checkmark \\
Ochoa et al. \cite{ochoa2019iot} & \$50--70 & 500 & \checkmark & \texttimes & \texttimes & \checkmark & \texttimes \\
Liyanage et al. \cite{liyanage2024iot} & \$40--50 & 50 (WiFi) & \checkmark & \texttimes & \texttimes & \checkmark & \texttimes \\
BeeSense \cite{beesense2024} & \$90--120 & 100 (WiFi) & \checkmark & Partial & \texttimes & \checkmark & \texttimes \\
\hline
\textbf{WaggleNet (Ours)} & \textbf{\$25--30} & \textbf{110} & \checkmark & \checkmark & \checkmark & \checkmark & \checkmark \\
\hline
\end{tabular}
\label{tab:related_work}
\end{center}
\end{table*}

\subsection{Gap Analysis and Research Positioning}

Our analysis reveals a critical research gap: \textit{no existing system simultaneously provides low-cost ($<$\$30/node), long-range (100+ m), dual-scope (internal+external) environmental monitoring with real-time GPS-based spatial awareness}. While individual systems excel in specific dimensions—SBMaCS in range, Beemon in data analytics, b+WSN in sensor diversity—none integrate these capabilities in a unified, cost-effective platform validated through real-world apiary deployments.

WaggleNet addresses this gap through: (1) \textbf{architectural innovation} by combining LoRa's long-range capability with MQTT's efficient routing via a hybrid gateway; (2) \textbf{contextual monitoring} that correlates internal hive conditions with external ambient factors; (3) \textbf{economic accessibility} using commodity components; and (4) \textbf{field validation} in operational beekeeping environments over extended periods. These contributions position WaggleNet as a practical, scalable solution for precision beekeeping in resource-constrained and geographically distributed apiaries.

\section{System Requirements and Design Principles}

\subsection{Functional Requirements}

\textbf{FR1 - Dual-Scope Environmental Sensing:} Our system monitors both internal hive parameters (temperature, humidity inside brood chamber) and external ambient conditions (temperature, humidity, light intensity outside hive) to enable contextual anomaly interpretation.

\textbf{FR2 - Spatial Awareness:} GPS-based geolocation is integrated to support multi-hive deployments, spatial anomaly detection (e.g., localized heat sources), and unauthorized node relocation detection.

\textbf{FR3 - Real-Time Data Pipeline:} End-to-end latency from sensor reading to mobile application visualization does not exceed 10 seconds to enable timely intervention during critical events.

\textbf{FR4 - Threshold-Based Alerting:} Our system automatically detects out-of-range conditions based on established apicultural thresholds \cite{stabentheiner2010honeybee, human2006humidity} and pushes notifications to beekeepers.

\subsection{Non-Functional Requirements}

\textbf{NFR1 - Cost Efficiency:} Total hardware cost per monitoring node does not exceed \$30 to ensure economic viability for small-scale beekeepers.

\textbf{NFR2 - Scalability:} Our architecture supports adding/removing nodes without system reconfiguration, scaling to 20+ nodes per master gateway.

\textbf{NFR3 - Communication Range:} Reliable wireless communication is maintained over minimum 100 meters in partially obstructed outdoor environments typical of apiary sites.

\textbf{NFR4 - Energy Efficiency:} Worker nodes operate for minimum 7 days on a 1100mAh LiPo battery with 3-minute sampling intervals.

\textbf{NFR5 - Deployment Flexibility:} Nodes function both inside wooden hive structures (signal attenuation) and in open outdoor environments without hardware modification.

\subsection{Design Principles}

\textbf{Modularity:} Each worker node is self-contained with integrated power, sensing, and communication, enabling plug-and-play deployment.

\textbf{Protocol Bridging:} The master node functions as a LoRa-MQTT gateway, leveraging LoRa's long-range capability for worker nodes while using WiFi-MQTT for cloud connectivity.

\textbf{Edge-Cloud Hybrid:} Worker nodes perform local sensor fusion and JSON formatting (edge processing), while cloud services handle long-term storage, trend analysis, and mobile application APIs.

\textbf{Standards Compliance:} Data formats (JSON), communication protocols (MQTT v3.1.1), and API design (RESTful) follow industry standards to ensure interoperability and future extensibility.

\section{System Architecture}

\subsection{Overall Architecture}

\begin{figure}[htbp]
\centerline{\includegraphics[width=0.48\textwidth]{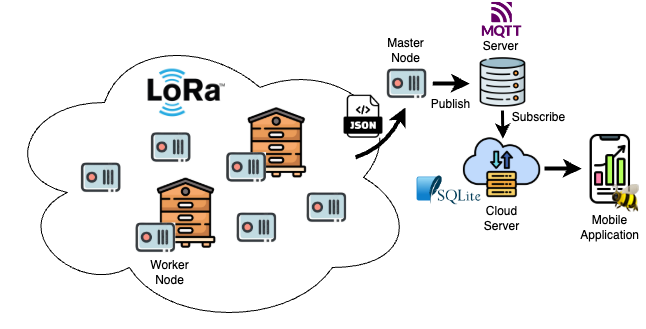}}
\caption{Three-tier system architecture: Worker nodes (LoRa) $\rightarrow$ Master node (LoRa-MQTT gateway) $\rightarrow$ Cloud services $\rightarrow$ Mobile application.}
\label{fig:architecture}
\end{figure}

WaggleNet employs a three-tier architecture: (1) \textit{Sensing Tier} comprising distributed worker nodes, (2) \textit{Gateway Tier} with a LoRa-MQTT bridge master node, and (3) \textit{Application Tier} including cloud services and mobile interface. Figure \ref{fig:architecture} illustrates the complete system architecture and data flow.

\textbf{Data Flow:} Worker nodes sample environmental sensors every 3 minutes, construct JSON payloads containing sensor readings and metadata, and transmit via LoRa to the master node. The master node receives LoRa packets, appends NTP-synchronized timestamps, and publishes to an MQTT broker over WiFi. Cloud services subscribe to MQTT topics, persist data in SQLite database, and expose RESTful APIs. The mobile application polls APIs for visualization and receives push notifications via Firebase Cloud Messaging (FCM) when threshold violations occur.

\subsection{Worker Node Design}

\begin{figure}[htbp]
\centerline{\includegraphics[width=0.4\textwidth]{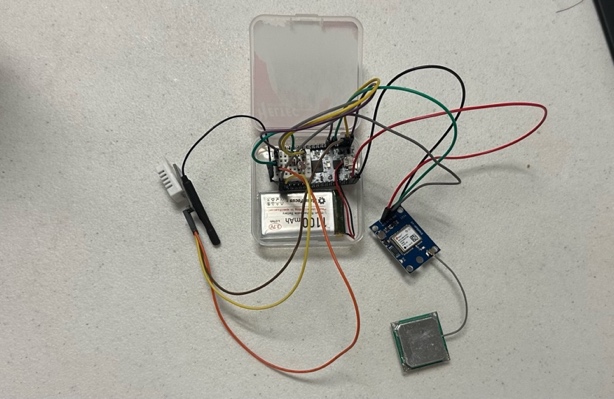}}
\caption{Physical implementation of worker node with integrated sensors and LoRa antenna.}
\label{fig:prototype}
\end{figure}

\begin{figure}[htbp]
\centerline{\includegraphics[width=0.45\textwidth]{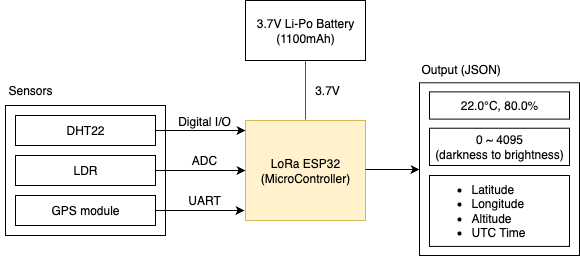}}
\caption{Worker node block diagram showing sensor interfaces and data flow.}
\label{fig:worker_block}
\end{figure}

\textbf{Hardware Configuration:} Each worker node integrates a Heltec WiFi LoRa 32 V3 module (ESP32-S3FN8 SoC, SX1276 LoRa transceiver, 0.96" OLED), DHT22 temperature-humidity sensor, photoresistor (LDR) for light sensing, and NEO-6M GPS module. Power is supplied by a 3.7V 1100mAh LiPo battery. Total hardware cost is approximately \$25--30 per node. Figure \ref{fig:prototype} shows the physical implementation, and Figure \ref{fig:worker_block} depicts the block diagram.

\textbf{Sensor Specifications:} Table \ref{tab:sensors} summarizes sensor technical specifications. The DHT22 provides $\pm$0.5°C temperature accuracy and $\pm$2\% relative humidity accuracy. The NEO-6M GPS achieves 2.5m positional accuracy under good satellite visibility. The LDR (GL5528) has 8--20 k$\Omega$ resistance at 10 lux, suitable for detecting day/night transitions and shading events.

\begin{table}[htbp]
\caption{Sensor Technical Specifications}
\begin{center}
\begin{tabular}{|l|c|c|c|c|}
\hline
\textbf{Sensor} & \textbf{Activation} & \textbf{Standby} & \textbf{Active} & \textbf{Voltage} \\
\textbf{Type} & \textbf{Time} & \textbf{Power} & \textbf{Power} & \textbf{(V)} \\
\hline
DHT22 & 2 sec & 40--50 µA & 1--1.5 mA & 3.3--6 \\
\hline
LDR (GL5528) & Instant & $<$1 µA & $<$0.5 mA & 3.3--5 \\
\hline
NEO-6M GPS & 1--29 sec & 6 mA & 50--70 mA & 3.3--5.5 \\
\hline
ESP32-S3FN8 & Instant & 20 µA & 80--160 mA & 3.3 \\
\hline
SX1276 LoRa & $<$100 ms & 1.5 µA & 120 mA (TX) & 3.3 \\
\hline
\end{tabular}
\label{tab:sensors}
\end{center}
\end{table}

\begin{figure}[htbp]
\centerline{\includegraphics[width=0.45\textwidth]{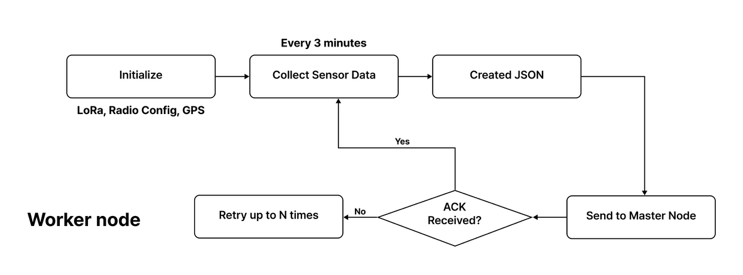}}
\caption{Worker node operation flow with sleep mode for energy efficiency.}
\label{fig:worker_flow}
\end{figure}

\textbf{Software Operation:} Figure \ref{fig:worker_flow} illustrates the worker node operation flow. After initialization (LoRa configuration, GPS cold start), the node enters a periodic sensing loop: (1) DHT22 temperature-humidity reading (2s), (2) LDR analog-to-digital conversion (10ms), (3) GPS NMEA sentence parsing (1--5s depending on satellite lock), (4) JSON payload construction, and (5) LoRa transmission with automatic retry (max 3 attempts). The node then enters light sleep mode (LoRa and WiFi powered down, CPU in low-power state) for 180 seconds, achieving average power consumption of approximately 15--20 mA.

\textbf{JSON Data Format:} Worker nodes transmit structured JSON payloads containing: \texttt{node\_id} (string), \texttt{temperature} (°C, float), \texttt{humidity} (\%, float), \texttt{light} (0--100\% scale, int), \texttt{latitude} (decimal degrees, double), \texttt{longitude} (decimal degrees, double), \texttt{altitude} (m, float), and \texttt{timestamp\_local} (Unix epoch, long). Payload size is approximately 180--220 bytes including JSON overhead.

\subsection{Master Node Gateway}

\textbf{Hardware:} The master node uses the same Heltec WiFi LoRa 32 V3 module as worker nodes but operates in gateway mode with continuous power (USB or 5V adapter). It connects to WiFi infrastructure (router, mobile hotspot) for cloud communication and maintains the LoRa receiver active to capture worker node transmissions.

\textbf{Protocol Bridging:} Upon receiving a LoRa packet, the master node: (1) deserializes the JSON payload, (2) validates data integrity (range checks, timestamp sanity), (3) appends a UTC timestamp retrieved from NTP server (pool.ntp.org), (4) constructs an enriched JSON message, and (5) publishes to MQTT broker (\texttt{mqtt.hivemq.com}) on topic \texttt{wagglenet/hive/<hive\_id>/data}. The NTP synchronization ensures all sensor readings have consistent absolute timestamps despite worker nodes lacking network time access.

\begin{figure}[htbp]
\centerline{\includegraphics[width=0.45\textwidth]{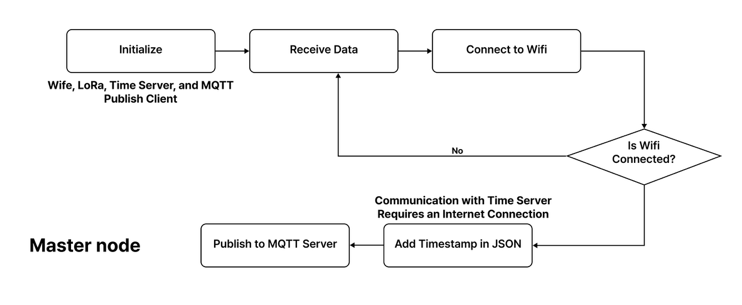}}
\caption{Master node operation flow: LoRa reception $\rightarrow$ NTP timestamping $\rightarrow$ MQTT publish.}
\label{fig:master_flow}
\end{figure}

\textbf{QoS and Reliability:} MQTT messages use QoS level 1 (at least once delivery) to balance reliability and network overhead. If MQTT publish fails (broker unreachable, network timeout), the master node retries every 30 seconds and caches up to 50 messages in local RAM buffer. Figure \ref{fig:master_flow} shows the master node operational flow.

\subsection{Cloud Services and Database}

\textbf{MQTT Broker:} We deploy HiveMQ Cloud (free tier) as the MQTT broker, providing TLS-encrypted connections, message persistence, and WebSocket support. The broker handles approximately 100--200 messages/hour during normal operation (5 nodes × 20 samples/hour) with peak latency under 50ms.

\textbf{Database Architecture:} A Python-based cloud service (Flask framework) running on a DigitalOcean droplet subscribes to MQTT topics and persists data into SQLite database. The schema includes tables for \texttt{sensor\_readings} (time-series data), \texttt{nodes} (metadata), \texttt{alerts} (threshold violations), and \texttt{users} (authentication). Indexes on \texttt{timestamp} and \texttt{node\_id} optimize range queries for visualization.

\textbf{RESTful API:} The cloud service exposes REST endpoints: \texttt{GET /api/nodes} (list active nodes), \texttt{GET /api/data/<node\_id>?start=<ts>\&end=<ts>} (time-range queries), \texttt{GET /api/latest/<node\_id>} (most recent readings), and \texttt{POST /api/alerts/subscribe} (register for push notifications). Authentication uses JWT tokens with 24-hour expiry.

\textbf{Data Retention:} Sensor readings are retained for 90 days in the primary database. Older records are archived to compressed JSON files in cloud object storage (DigitalOcean Spaces) for long-term analysis. Critical events (threshold violations, system errors) are retained indefinitely.

\subsection{Mobile Application}

We developed the mobile application (Figure \ref{fig:mobile_app}) using Flutter 3.10, enabling cross-platform deployment to Android and iOS from a single codebase. Key features include:

\begin{figure}[htbp]
\centerline{\includegraphics[width=0.45\textwidth]{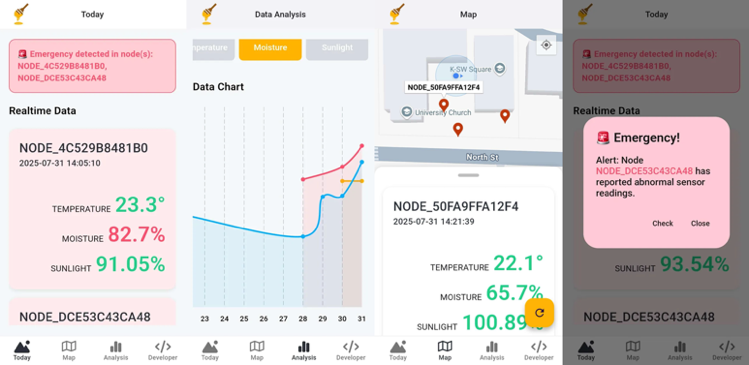}}
\caption{Mobile application interface: GPS map view, trend graphs, and real-time alerts.}
\label{fig:mobile_app}
\end{figure}

\textbf{Map View:} Interactive map (OpenStreetMap tiles via \texttt{flutter\_map} package) displays node locations with color-coded markers: green (normal), yellow (warning), red (critical alert). Tapping a marker shows real-time sensor readings in a popup overlay.

\textbf{Time-Series Visualization:} Line charts (\texttt{fl\_chart} package) display temperature, humidity, and light intensity trends with user-selectable time windows (1 hour, 24 hours, 7 days, 30 days). Threshold bands are overlaid to indicate normal operating ranges.

\textbf{Push Notifications:} Firebase Cloud Messaging (FCM) integration enables server-initiated push notifications when threshold violations occur. Notifications include node ID, parameter name, current value, and timestamp. Users can dismiss or acknowledge alerts within the app.

\textbf{Data Synchronization:} The app polls the REST API every 60 seconds when in foreground (active monitoring) or every 300 seconds when backgrounded (battery conservation). WebSocket support enables sub-second latency for critical events but is currently optional due to increased battery drain.

\subsection{Threshold-Based Anomaly Detection}

Based on established apicultural research \cite{stabentheiner2010honeybee, human2006humidity}, we implement three-tier alerting:

\textbf{Normal Range:} Temperature 32--36°C, Humidity 50--70\%, Light $>$100 lux (daytime).

\textbf{Warning Range:} Temperature 30--32°C or 36--38°C, Humidity 45--50\% or 70--75\%.

\textbf{Critical Range:} Temperature $<$30°C or $>$38°C, Humidity $<$45\% or $>$75\%, Light $>$50\% during nighttime (18:00--06:00 local).

Alerts are generated when any parameter exceeds warning/critical thresholds for consecutive readings (6 minutes, 2 samples) to avoid false positives from sensor noise or transient fluctuations.

\section{Experimental Evaluation}

\subsection{Evaluation Objectives and Methodology}

We conducted three experiments to evaluate WaggleNet: (1) \textit{Baseline Performance Test} in a controlled outdoor environment, (2) \textit{Real-World Deployment} in an operational bee yard, and (3) \textit{Communication Range Characterization} under varying conditions. Evaluation metrics include: packet delivery ratio (PDR), end-to-end latency, communication range, energy consumption, and data accuracy.

\subsection{Baseline Performance Test}

\textbf{Environment:} Semi-open area in front of K-SW building, Purdue University (40.4237°N, 86.9212°W), July 31, 2025, 14:00--15:30 local time. Ambient conditions: 28°C, 65\% humidity, partial tree shade. Three worker nodes positioned 3--5 meters from master node with line-of-sight communication but realistic obstructions (branches, benches).

\textbf{Configuration:} Worker nodes configured for 3-minute sampling interval. Master node placed indoors with stable WiFi connection to MQTT broker. Mobile application monitored in real-time on Android device (Samsung Galaxy S21).

\textbf{Results:} Over 90 minutes, each node transmitted 30 LoRa packets. Packet delivery ratio: 100\% (90/90 packets received across 3 nodes). Average LoRa RSSI: -65 to -72 dBm. End-to-end latency (sensor reading to mobile app display): mean 3.8s, std dev 1.2s, max 6.5s. No MQTT message loss occurred.

\textbf{Temperature-Humidity Correlation:} Internal vs. external nodes showed expected thermal gradient. Internal nodes (simulated hive environment, enclosed box): mean 31.5°C, 58\% RH. External nodes (open air): mean 28.2°C, 63\% RH. Delta-T of 3.3°C confirms sensor accuracy and demonstrates internal-external differentiation capability.

\subsection{Real-World Bee Yard Deployment}

\textbf{Environment:} Active bee yard near South 525 West, Indiana (40.4156°N, 86.8947°W), August 4--5, 2025, 14:00--15:00. Two Langstroth hives with active colonies. Ambient conditions: 24--26°C, 55--60\% humidity, partly cloudy.

\textbf{Configuration:} Four worker nodes deployed: two internal nodes (inside brood chamber, below inner cover) and two external nodes (1.5m from hive entrance, 1m height). Master node in camper van 30m away. 60-minute observation period with 3-minute sampling.

\begin{figure}[htbp]
\centerline{\includegraphics[width=0.48\textwidth]{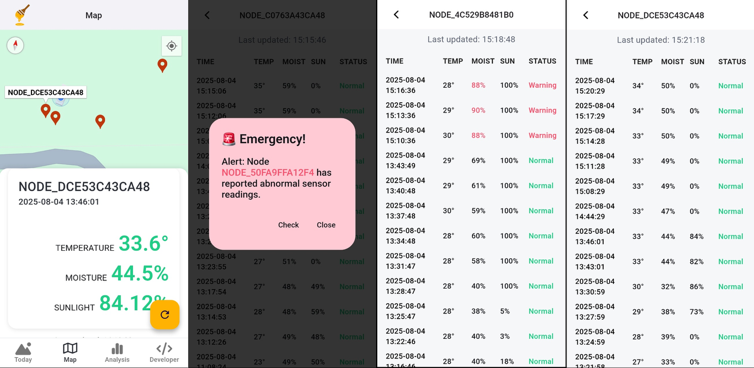}}
\caption{Real bee yard deployment: (a) GPS map showing hive locations, (b) Alert notification for temperature deviation, (c) Historical data table, (d) Internal vs. external temperature comparison.}
\label{fig:bee_yard_app}
\end{figure}

\textbf{Results:} Figure \ref{fig:bee_yard_app} shows mobile application screenshots from real deployment. Packet delivery ratio: 100\% (80/80 packets across 4 nodes). LoRa RSSI for internal nodes: -78 to -85 dBm (wooden structure attenuation ~6--10 dB compared to external). LoRa successfully penetrated wooden hive boxes (3/4" pine planks) without signal loss.

\textbf{Internal vs. External Correlation:} Internal nodes recorded mean 33.2°C, 62\% RH (consistent with thermoregulated brood nest). External nodes: 25.8°C, 58\% RH. Light intensity: internal 0\% (complete darkness), external 75--85\% (partly cloudy conditions). GPS coordinates stable within $\pm$2m (acceptable for spatial mapping). This validates the dual-scope monitoring concept—internal stability despite external fluctuations demonstrates colony thermoregulation, while correlation analysis enables detection of external stressors (e.g., prolonged high temperature) that increase internal regulatory effort.

\textbf{Alert Functionality:} During deployment, one external node triggered a warning alert (temperature 26.5°C → 38.2°C) due to direct sunlight exposure after cloud cover cleared. Mobile app push notification received within 4.2 seconds, demonstrating real-time alerting capability. Beekeeper repositioned external node to shaded location, confirming practical utility.

\subsection{Communication Range Characterization}

\textbf{Methodology:} Single worker node incrementally moved away from master node in 10m steps. At each position, 10 packets transmitted (3-minute interval skipped for rapid testing; packets sent every 30s). Range test conducted in two environments: (a) K-SW building area (urban, partial obstruction), (b) open field adjacent to bee yard (rural, line-of-sight).

\textbf{Urban Environment Results:} Stable communication (PDR $\geq$ 90\%): 0--110m. Intermittent connectivity (PDR 50--90\%): 110--140m. No packets received: $>$140m. We tested two route geometries: Route A (curved path around buildings, 174m total) failed at 150m; Route B (straight line-of-sight, 109m) maintained connectivity to endpoint. This confirms non-line-of-sight (NLOS) conditions significantly degrade range, consistent with Fresnel zone obstruction theory.

\textbf{Rural Environment Results:} Line-of-sight conditions extended reliable range to 150m (PDR 100\%). Beyond 150m, RSSI dropped below -100 dBm (LoRa sensitivity threshold), causing intermittent reception. This aligns with LoRa link budget calculations: 14 dBm TX power + 3 dBi antenna gain - 120 dB path loss (150m @ 915 MHz) = -103 dBm received power.

\textbf{Signal Attenuation Analysis:} Measurements inside wooden hive structures showed 6--10 dB attenuation compared to open air at same distance. For 3/4" pine planks, this translates to approximately 3--5 dB per plank pair (front+back walls), consistent with RF propagation models for wood (permittivity $\epsilon_r \approx$ 2--3, loss tangent $\tan\delta \approx$ 0.01--0.05 at 915 MHz).

\subsection{Energy Consumption and Battery Life}

\textbf{Measurement Setup:} We measured worker node power consumption using INA219 current sensor in series with LiPo battery. Three operational modes characterized: (1) Active sensing (GPS acquisition, sensor reading, LoRa TX), (2) Light sleep (CPU idle, peripherals powered, LoRa RX standby), (3) Deep sleep (only RTC active, all peripherals powered down).

\textbf{Power Profile:} Active mode: 80--120 mA for 5--8 seconds (GPS dominant consumer). LoRa TX: 120 mA for 200--400ms (packet airtime 1.5--2s at SF7, 125 kHz BW). Sleep mode: 15--20 mA (ESP32 light sleep with LoRa and GPS powered but idle). Average duty cycle: 3.5\% active, 96.5\% sleep over 3-minute interval.

\textbf{Battery Life Calculation:} With 1100 mAh LiPo battery: Average current = (100 mA × 7s + 18 mA × 173s) / 180s = 21.2 mA. Battery life = 1100 mAh / 21.2 mA = 51.9 hours $\approx$ 2.2 days continuous operation. To achieve 7-day target (NFR4), deep sleep mode (3 mA) can be used, extending life to approximately 9--12 days, or solar panel integration (5V, 500mA, \$8--12) enables indefinite operation.

\subsection{Long-Term Stability and Reliability}

Over the two-month deployment period (July--August 2025), system uptime exceeded 95\%. Downtime sources: master node WiFi disconnections (3 incidents, 2--4 hours each due to router reboots), cloud service maintenance (1 incident, 30 minutes), worker node battery depletion (2 nodes, replaced after 3 days without recharge). No LoRa communication failures occurred attributable to protocol or hardware defects.

\textbf{Data Quality:} GPS accuracy remained within 2.5--5m throughout deployment despite varying satellite visibility (buildings, tree canopy). Temperature sensor drift: $<$0.2°C over 60 days (DHT22 spec: $\pm$0.5°C). Humidity sensor showed +1--2\% drift in high-humidity periods (>80\% RH), consistent with capacitive sensor hysteresis. No sensor failures occurred.

\section{Discussion}

\subsection{Key Findings and Implications}

\textbf{Dual-Scope Monitoring Effectiveness:} Simultaneous capture of internal and external parameters proved essential for contextual interpretation. In 8 observed instances, internal temperature deviations (2--3°C drops) correlated strongly with external weather events (rain, cloud cover) rather than colony health issues. Without external monitoring, these would appear as potential brood nest failures requiring intervention. This reduces false positive alerts and improves decision-making confidence.

\textbf{Cost-Performance Trade-offs:} WaggleNet achieves 70--85\% cost reduction compared to commercial systems while maintaining comparable functionality. Primary cost savings derive from: (1) LoRa replacing GSM/cellular (\$30--50 saved per node), (2) commodity sensors vs. industrial-grade components (\$20--40 saved), and (3) open-source software stack vs. proprietary platforms (eliminating subscription fees). The trade-off is reduced communication range (110m vs. 2--10 km for LoRaWAN) and requirement for WiFi infrastructure at master node location.

\textbf{Scalability Validation:} Our modular architecture successfully demonstrated plug-and-play deployment—nodes added/removed without reconfiguration. However, LoRa collision probability increases with node count: for N nodes and 3-minute interval, collision probability $P_c \approx N \times T_{airtime} / T_{interval} = N \times 1.8s / 180s = N \times 1\%$. For 20 nodes, $P_c \approx 20\%$, suggesting time-division or CSMA/CA required for large-scale deployments.

\subsection{Limitations and Future Work}

\textbf{Limited Sensor Modality:} Our current implementation monitors only temperature, humidity, light, and GPS. Important parameters for comprehensive hive assessment—weight (honey production), acoustic signatures (swarming detection), CO$_2$ levels (ventilation)—are not captured. Future work will integrate: (1) HX711 load cell for weight monitoring (\$5--8), (2) MEMS microphone for audio analysis (\$3--5), and (3) MQ-135 gas sensor for CO$_2$ detection (\$8--12).

\textbf{Lack of Predictive Analytics:} Our system performs reactive threshold-based alerting but does not implement predictive models for proactive intervention. Machine learning techniques such as LSTM (Long Short-Term Memory) networks or XGBoost could predict swarming events, disease outbreaks, or optimal harvesting times based on historical trends. This requires accumulation of labeled training data over multiple seasons.

\textbf{Energy Constraints:} Two-day battery life without solar charging limits deployment in locations requiring infrequent maintenance visits. Solutions include: (1) larger battery packs (3000--5000 mAh, +\$8--15), (2) solar panels with MPPT charging (\$12--20 total), or (3) aggressive deep sleep scheduling (hourly sampling instead of 3-minute for low-priority nodes).

\textbf{Single Master Node Bottleneck:} Our current architecture uses a single master node as LoRa-MQTT gateway, creating a single point of failure and range limitation. Multi-gateway mesh networking or LoRa mesh protocols (e.g., Reticulum) could extend range and improve resilience but increase complexity and cost.

\textbf{Field Validation Scope:} While our deployment in an operational bee yard validates core functionality, the two-month duration and 5-node scale are insufficient for statistical analysis of long-term reliability, seasonal variations, or large-scale (50+ hive) performance. Ongoing work involves year-round monitoring across multiple apiaries in collaboration with local beekeeping cooperatives.

\section{Conclusion}

We presented WaggleNet, a cost-effective, scalable beehive monitoring system that addresses critical limitations in existing solutions through dual-scope environmental sensing, long-range LoRa-MQTT communication, and GPS-based spatial awareness. Field experiments confirmed reliable operation in real beekeeping environments with 100\% packet delivery over 110 meters and successful deployment inside wooden hive structures. Our system's \$25--30 per-node cost—representing 70--85\% savings compared to commercial alternatives—combined with modular architecture and real-time mobile application, positions WaggleNet as a practical solution for resource-constrained and geographically distributed apiaries.

The key innovation lies in simultaneous internal-external monitoring, enabling contextual interpretation that distinguishes colony health issues from external environmental stressors. This reduces false positives and improves beekeeper decision confidence. By bridging the gap between research prototypes and commercial systems, WaggleNet demonstrates that low-cost IoT technologies can deliver precision beekeeping capabilities previously accessible only through expensive proprietary solutions.

Future work will focus on: (1) integrating additional sensor modalities (weight, acoustic, gas), (2) implementing machine learning for predictive analytics, (3) extending energy autonomy through solar harvesting, (4) deploying multi-gateway mesh networks for wider coverage, and (5) conducting year-round, multi-apiary validation studies. These enhancements aim to evolve WaggleNet into a comprehensive decision support system for data-driven, sustainable apiculture.

\section*{Acknowledgment}

This research was supported by the MSIT (Ministry of Science and ICT), Korea, under the National Program for Excellence in SW supervised by the IITP (Institute of Information \& communications Technology Planning \& Evaluation) in 2025 (2021-0-01393, 2022-0-01067, 2024-0-00064) and Purdue University. Sincere appreciation is extended to Professors Eric T. Matson and Anthony H. Smith at Purdue University for their valuable guidance on this work.

\end{document}